\documentclass[twocolumn,secnumarabic,amssymb, nobibnotes, aps, prd, floatfix,nofootinbib]{revtex4-2}

\usepackage{graphicx}
\usepackage{amssymb}
\usepackage{mathtools}
\usepackage{amsmath}
\usepackage{xcolor}
\usepackage{bm}
\usepackage{fancyhdr}
\usepackage{float}
\usepackage{braket}
\usepackage{booktabs}
\usepackage{adjustbox}
\usepackage{soul} 

\DeclareMathAlphabet{\mathbbold}{U}{bbold}{m}{n} 

\makeatletter
\newcommand{\myfnsymbol}[1]{%
  \expandafter\@myfnsymbol\csname c@#1\endcsname
}
\newcommand{\@myfnsymbol}[1]{%
  \ifcase #1
  \or \TextOrMath{\textasteriskcentered}{*}
  \or $\dag$
  \or $\ddag$
  \or $\P$
  \fi
}
\newcommand{\equalcontributor}{\@myfnsymbol{1}}
\newcommand{\affiliationA}{\@myfnsymbol{2}}
\newcommand{\affiliationB}{\@myfnsymbol{3}}
\newcommand{\affiliationC}{\@myfnsymbol{4}}
\makeatother

\begin{document}

\title{Spectrally shaped and pulse-by-pulse multiplexed \\ multimode squeezed states of light}%

\author{Tiphaine Kouadou\textsuperscript{\equalcontributor,\affiliationA}}
\affiliation{Laboratoire Kastler Brossel, Sorbonne Universit\'e, CNRS,ENS-PSL Research University, Coll\`ege de France\\4 place Jussieu, F-75252, Paris, France}
\author{Francesca Sansavini\textsuperscript{\equalcontributor,\affiliationB}}%
\affiliation{Laboratoire Kastler Brossel, Sorbonne Universit\'e, CNRS,ENS-PSL Research University, Coll\`ege de France\\4 place Jussieu, F-75252, Paris, France}
\author{Matthieu Ansquer}%
\affiliation{Laboratoire Kastler Brossel, Sorbonne Universit\'e, CNRS,ENS-PSL Research University, Coll\`ege de France\\4 place Jussieu, F-75252, Paris, France}
\author{Johan Henaff}%
\affiliation{Laboratoire Kastler Brossel, Sorbonne Universit\'e, CNRS,ENS-PSL Research University, Coll\`ege de France\\4 place Jussieu, F-75252, Paris, France}
\author{Nicolas Treps}%
\affiliation{Laboratoire Kastler Brossel, Sorbonne Universit\'e, CNRS,ENS-PSL Research University, Coll\`ege de France\\4 place Jussieu, F-75252, Paris, France}
\author{Valentina Parigi \textsuperscript{\affiliationC}}%
\affiliation{Laboratoire Kastler Brossel, Sorbonne Universit\'e, CNRS,ENS-PSL Research University, Coll\`ege de France\\4 place Jussieu, F-75252, Paris, France}

\begin{abstract}
\noindent 
Spectral- and time- multiplexing are currently explored to generate large multipartite quantum states of light for quantum technologies. In the continuous variable approach, the deterministic generation of large entangled states demands the generation of a large number of squeezed modes. Here, we demonstrate the simultaneous generation of 21 squeezed spectral modes at 156 MHz.  We exploit the full repetition rate and the ultrafast shaping of a femtosecond light source to combine, for the first time, frequency- and time- multiplexing in multimode squeezing. This paves the way to the implementation of multipartite entangled states that are both scalable and fully reconfigurable. 
 
\end{abstract}

\renewcommand{\thefootnote}{\myfnsymbol{footnote}}
\maketitle
\footnotetext[1]{These authors contributed equally to this work}%
\footnotetext[2]{tiphaine@illinois.edu. Current address: Department of Physics, University of Illinois at Urbana-Champaign, Urbana, IL 61801, USA}%
\footnotetext[3]{francesca.sansavini@lkb.upmc.fr}%
\footnotetext[4]{valentina.parigi@lkb.upmc.fr}%


\section{Introduction}

Multimode quantum light fields enable  continuous-variable (CV)-based quantum information technologies. In particular, light fields squeezed in different spatial, spectral and temporal shapes are a versatile and promising resource for measurement-based quantum computing (MBQC) \cite{Menicucci2006, Gu09}, quantum-enhanced sensing \cite{Zhang2021,Pinel2012, Gessner2018}, multiparty communication protocols \cite{Cai17,Arzani19}, as well as simulation \cite{Nokkala18a} and quantum-enhanced machine learning protocols \cite{Nokkala21}.

The leverage of the continuous-variable approach, that encodes quantum information in the amplitude and phase values of the light field, is the deterministic generation of multipartite entangled states using nonlinear optical processes. Cluster states, a necessary resource for MBQC \cite{Raussendorf2001,Raussendorf2003,Menicucci2006,Gu09}, are the largest entangled structures shaped from  spectral and temporal modes of squeezed light \cite{Chen14,Cai17,Yokoyama13,Asavanant19,Larsen19}.
Beyond the amount of generated squeezing, two crucial features of CV entangled quantum resources are the scalability, i.e., the number of entangled parties, and the reconfigurability, i.e., the capability to reshape the entanglement links at will. The CV entangled states are generated by mixing several squeezed optical modes using linear-optics operations or, more generally, via mode-basis changes. As expected the scalability depends on the number and the production rate of the of initial squeezed modes. Differently from the discrete variables case  \cite{Wein22,Istrati20,Schwartz16}, the number of the required squeezed modes also strongly depends on the shapes of the entangled structures  \cite{Gu09,Centrone21}. It is in then crucial to easily access basis changes to reshape the quantum links.

Here we demonstrate the generation of multimode squeezed states that are multiplexed both in the temporal and spectral degrees of freedom. While temporal multiplexing already enabled the generation of the largest CV-cluster states \cite{Asavanant19,Larsen19}, multimode squeezing in the spectral modes of a femtosecond source, enables the full reconfigurability of the entanglement and mode-selective non-gaussian operations via well-controlled ultrafast shaping techniques \cite{Cai17,Ra20}. Both ingredients are present in our experimental setup: the time multiplexing is reached by measuring squeezing in individual laser pulses  emitted at a repetition rate of 156 MHz, moreover 21 squeezed spectral modes are generated for each pulse. 

The quantum resource is obtained via spontaneous parametric down conversion (SPDC) in waveguides 
in a single-pass configuration. Nonlinear waveguides are promising platforms for the controlled generation of both single- and  multimode quantum states \cite{Eckstein11,Harder16,Brecht15,Ansari18}.
Indeed, they have large nonlinearities and  can be engineered to tailor the modal decomposition of the generated quantum state \cite{Ansari18OP}.
In particular, PPKTP waveguides provide the suitable phase-matching conditions for modal engineering compatible with mode-selective femtosecond homodyne detection \cite{Roman2020}. Periodically poled waveguides have demonstrated fully integrated \cite{Kaiser16}, and large \cite{kashiwazaki20} single-mode squeezing, and in particular, PPKTP crystals have shown controlled squeezing generation in amplified schemes \cite{Akihito2021,Zhenju2018}. While this work focuses on the generation of a large number of squeezed spectral modes, several integrated platforms are tailored to limit the number of spectral and temporal modes involved, in order to increase the level of squeezing per mode \cite{Chen14,Yang2021,Madsen2022}.
In the temporal domain, state-of the-art experiments have been performed to generate cluster states  \cite{Asavanant19,Larsen19}, and boson sampling experiments \cite{Madsen2022}. 

\begin{figure*}
\includegraphics[width=\textwidth]{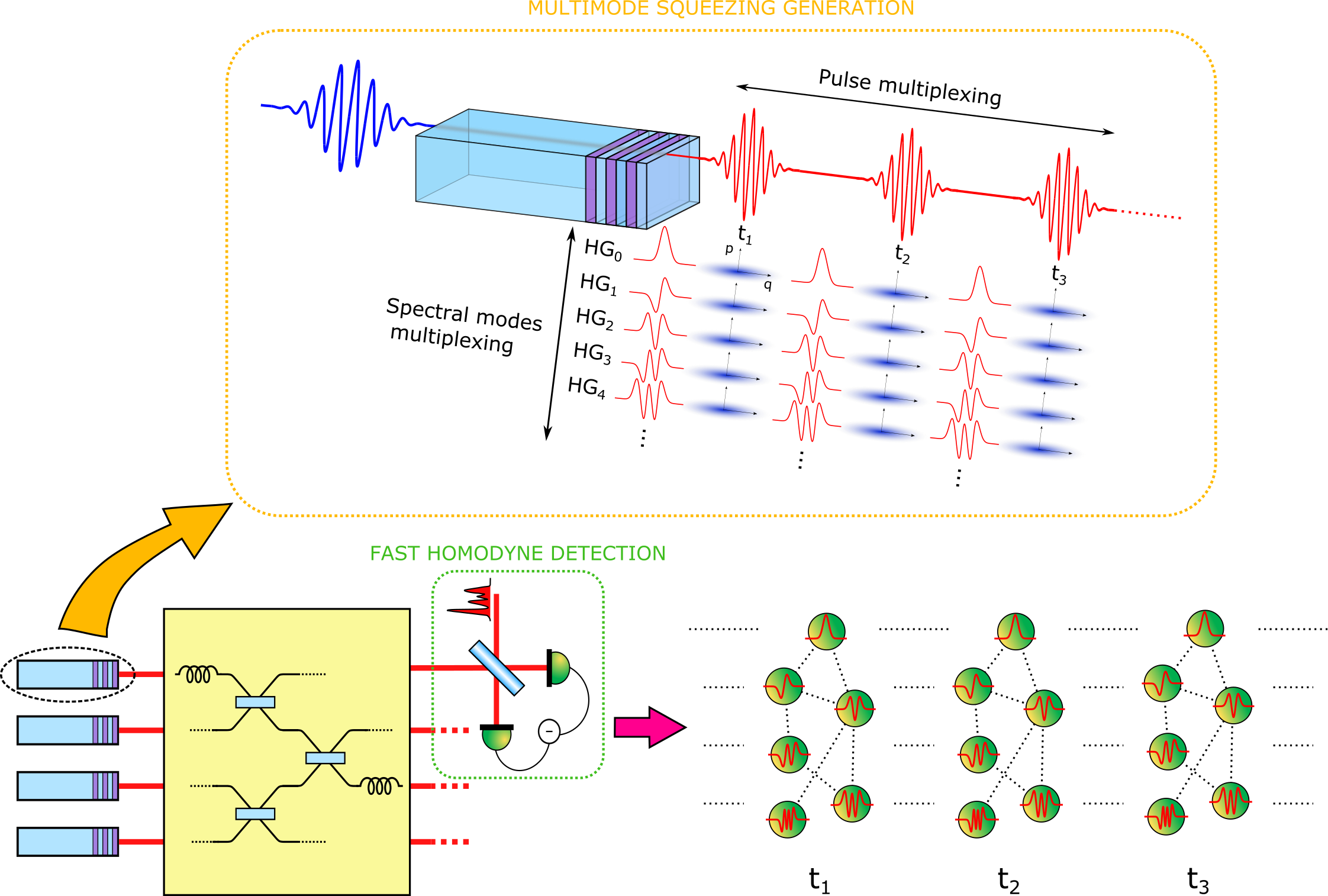}
\caption{Scheme of the scaling of the generated  quantum resource. Upper part in the dotted box: the generation of multimode squeezed states multiplexed both in time (rate of 156 MHz) and optical frequencies (21 Hermite Gauss modes, width of the first mode $\sim$ 18 nm) that is reported in this work  and measured via a spectrally mode-selective and fast homodyne detector (shown in the lower green dotted box). Lower part left: possible scaling by combining multiple waveguides and the spectrally mode-selective homodyne at the output with delay lines and beam-splitters. Lower part right: graphical representation of a 3-dimensional CV quantum network that can be generated by the scheme on the left with at least two wave-guides. The structure is composed by several entangled frequency modes  at given time and correlations between and arbitrary number of different temporal layers  . }
\label{scheme}
\end{figure*}

Here we multiplex , for the first time,  both in time at high repetition rate (156 MHz) and in reconfigurable spectral shapes (ultrafast shaping).  Moreover we show entanglement correlations between  optical-frequency bands.
The demonstrated resource, composed by a tailored non-linear waveguide and ultra fast mode-selective homodyne with large electronic bandwidth,  is the pivotal building block for scalable and reconfigurable networks as shown in Fig. \ref{scheme}.  By combining at least two waveguides \cite{Asavanant19,Larsen19} with delay lines and linear optics,  as shown in the lower part of  Fig. ~\ref{scheme}, the entanglement structure in the optical frequency \cite{Cai17} can be extended in the time (pulse-based) domain.

\section{Results}

\paragraph*{\textbf{Spectrally Multimode Squeezed Light}}
We use SPDC in a periodically poled Potassium Titanyl Phosphate (PPKTP) waveguide to generate spectrally multimode squeezed states of light  at the repetition rate of 156 MHz.  The train of  22-fs pulses, emitted by the light source, is converted via second harmonic generation (SHG) and is used to pump a 1-mm long PPKTP waveguide. The waveguide is designed for type 0 phase-matching corresponding to the largest nonlinear coefficient in KTP. While most multimode squeezed sources of light to date were obtained using resonant cavities \cite{Chen14,Roslund14,Cai17,Asavanant19,Larsen19}, we reach high pump intensity via 
 light confinement into a waveguide. This allows for the generation of trains of quadrature-squeezed pulses in a single passage.

In this framework, the Hamiltonian in the interaction picture  associated to a single pulse $m$ can be expressed in an eigenbasis of squeezed frequency modes \cite{Roslund14}:
\begin{equation}
\hat{H}_{\text{int}}^{m} =   \mathcal{K}  \, \sum_j \, \lambda_j \, \left( \hat{s}_j^{\dagger} \right)^2 \, + \, h.c.
\label{eq:squeezed_ham}
\end{equation}
where $\mathcal{K}$ is a constant that depends on the nonlinearity of the waveguide and the amplitude of the pump field, $\{ \hat{s}_j^\dagger \}$ is a set of creation operators associated with spectral-mode functions $S_j(\omega)$, that we call \textit{supermodes} and the $\{\lambda_j\}$ are related to the amount of squeezing in the spectral modes. In a collinear degenerate SPDC, where signal and idler fields are indistinguishable, the eigenmodes can be retrieved using a Bloch-Messiah reduction (equivalent to a Schmidt decomposition) \cite{Horoshko19}.


The wide bandwidth of the type-0 phase-matching in KTP, the pump spectral width and the length of the periodically poled region in the waveguide lead to the generation of numerous squeezed modes, as assessed by the estimated Schmidt parameter K = 98. We choose such a configuration to show the ability of the system to generate a high number of optical spectral modes, that can be independently measured via homodyne detection.

 We reveal  the amount of squeezing in multiple spectral modes by spectral shaping of the homodyne reference beam, named local oscillator (LO) \cite{Roslund14,Cai17}. After light detection, we obtain the variance of the spectral-mode quadrature by measuring the sideband intensity noise on a spectrum analyzer. The squeezed (anti-squeezed) quadrature displays noise below (above) the shot noise.
 
Table \ref{tab:sqz} shows the squeezing measurement results along with the corresponding spectral shape of the LO. We measured squeezing in 21 supermodes (up to HG$_{20}$) with 0.47 dB in the leading mode. The width of Hermite-Gauss functions is known to increase with their order $n$, therefore the number of supermodes that we can experimentally access is limited by the 40-nm wide spectrum of the LO. 
\begin{table}
\centering
\begin{tabular}{cccc}
\toprule
 \multicolumn{2}{c}{Mode}																			&Sqz (dB)						&Asqz (dB)	\\		
\midrule	
HG0		&\begin{adjustbox}{valign=c}\includegraphics[width=0.1\textwidth]{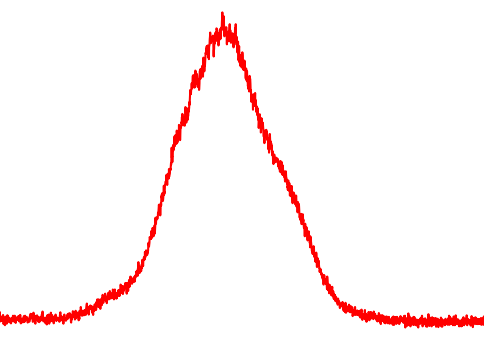} \end{adjustbox}		&-0.47				&0.55\\
HG1		&\begin{adjustbox}{valign=c}\includegraphics[width=0.1\textwidth]{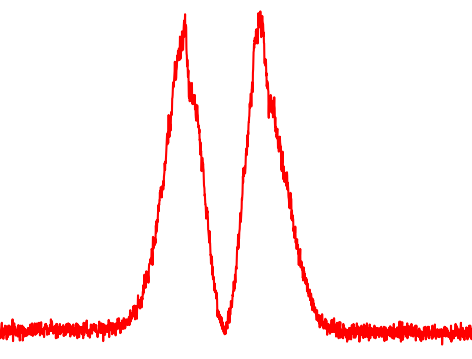}\end{adjustbox}		&-0.33				&0.42\\
HG2		&\begin{adjustbox}{valign=c}\includegraphics[width=0.1\textwidth]{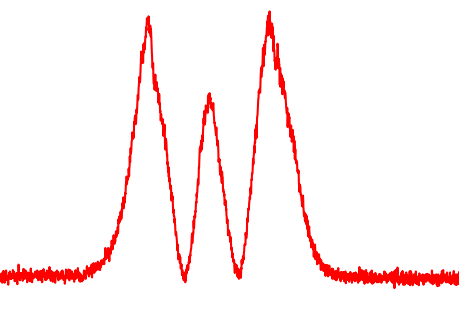}\end{adjustbox}		&-0.23				&0.35\\
HG3		&\begin{adjustbox}{valign=c}\includegraphics[width=0.1\textwidth]{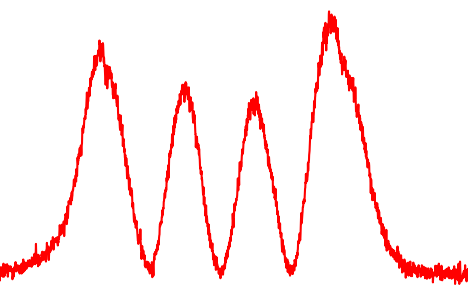}\end{adjustbox}		&-0.20				&0.28\\
HG4		&\begin{adjustbox}{valign=c}\includegraphics[width=0.1\textwidth]{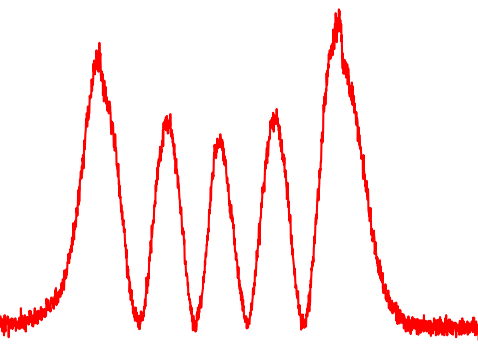}\end{adjustbox}		&-0.17				&0.30\\
HG5		&\begin{adjustbox}{valign=c}\includegraphics[width=0.1\textwidth]{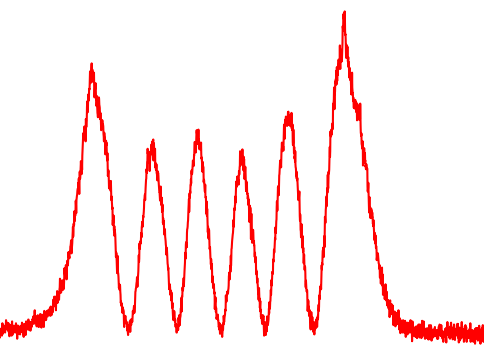}\end{adjustbox}		&-0.18				&0.30\\
HG6		&\begin{adjustbox}{valign=c}\includegraphics[width=0.1\textwidth]{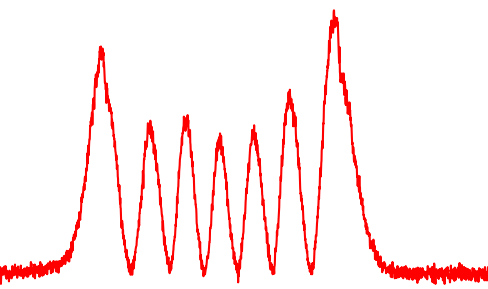}\end{adjustbox}		&-0.18				&0.27\\
HG7		&\begin{adjustbox}{valign=c}\includegraphics[width=0.1\textwidth]{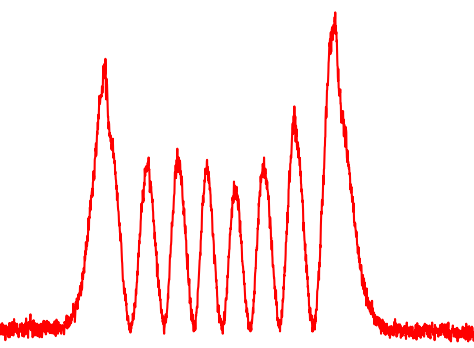}\end{adjustbox}		&-0.14				&0.26\\
HG8		&\begin{adjustbox}{valign=c}\includegraphics[width=0.1\textwidth]{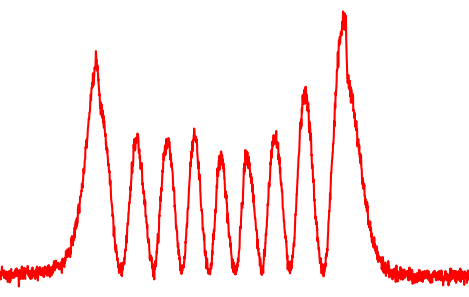}\end{adjustbox}		&-0.15				&0.26\\
\midrule
HG15	&\begin{adjustbox}{valign=c}\includegraphics[width=0.1\textwidth]{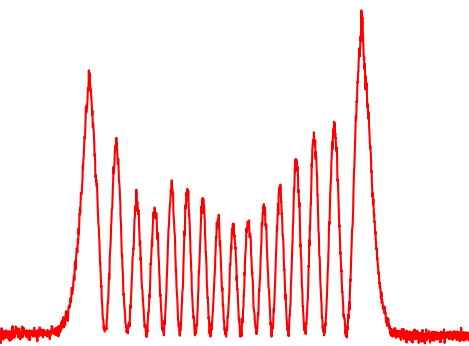}\end{adjustbox}		&-0.09				&0.17\\
\midrule
HG20	&\begin{adjustbox}{valign=c}\includegraphics[width=0.1\textwidth]{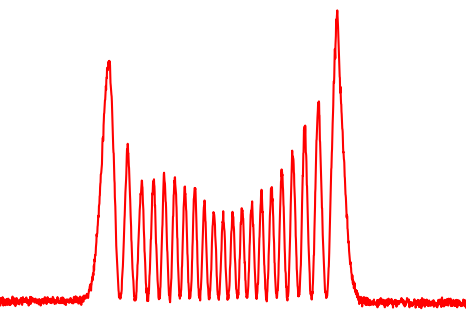}\end{adjustbox}		&-0.08				&0.16\\

\bottomrule
\end{tabular}
\caption{Level of squeezing (Sqz) and antisqueezing (Asqz) measured at a sideband frequency of 50 MHz in various Hermite-Gauss modes, along with the corresponding LO spectrum used for the measurement. The highest order mode measured is HG$_{20}$.}
\label{tab:sqz}
\end{table}
Since we chose a configuration with a large number of squeezed spectral modes, the squeezing per mode is expected to be low, despite the experiment being low-loss: there is in fact a trade-off between the number of generated squeezed modes and the squeezing level per mode. The high nonlinearity of KTP, the pump intensity  and the length of the poled region set the total amount of squeezing that can be produced, while its distribution into a given number of spectral modes depends on the phase-matching configuration, the spectral width of the pump and the  waveguide dimensions \cite{Roman2020, Dirmeier20}.

Importantly, while we have revealed the number of squeezed mode, this source can be reshaped via the adaptation of the measurement device into arbitrary entangled states, making it a promising resource for CV protocols \cite{Cai17}.


\paragraph*{\textbf{Simultaneous time and frequency multiplexing}}
To verify that the set of spectrally squeezed modes (Fig. \ref{scheme}) is produced at each pulse $m$ of the train, 
we characterize the signal with the temporal resolution  of the individual pulses. To do so, we have developed an homodyne detector with wide electronic bandwidth, a  cutoff frequency larger than the laser repetition rate 
and a large common mode rejection ratio (CMRR) \cite{Kumar12}. 



We acquire the homodyne signal
via a fast oscilloscope: one quadrature value is associated to a single pulse, by integrating the scope signal on the pulse time window. The latter has to be carefully adjusted to precisely discriminate between consecutive pulses. 
Details of the experimental protocols can be found in Methods.

\begin{figure}
\includegraphics[width=0.45\textwidth]{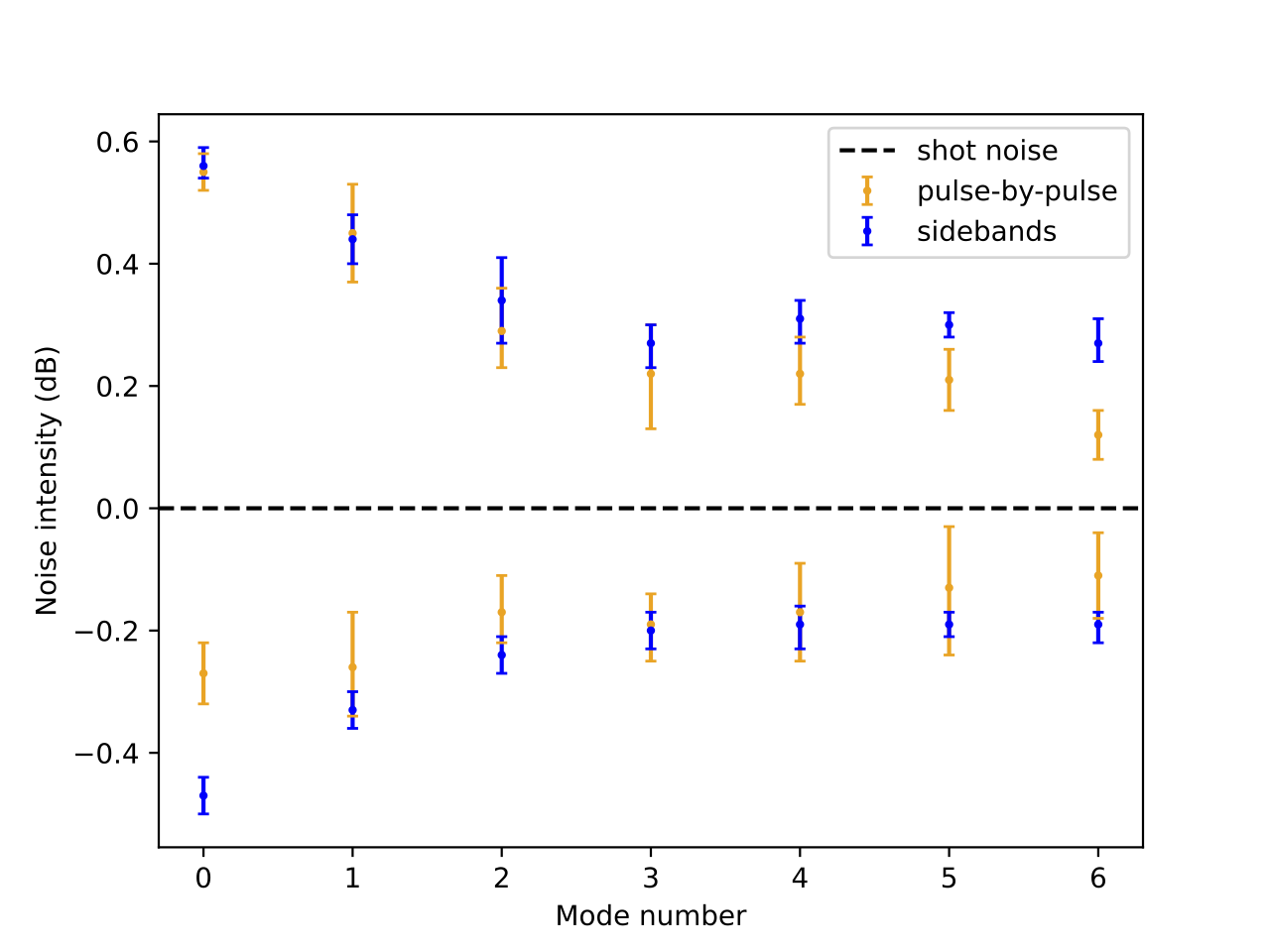}
\caption{Comparison of the sidebands and pulse-by-pulse squeezing and antisqueezing measurements for the first 6 HG supermodes. The two measurement sets share the same basis but different measurement techniques and devices: spectrum analyser for the sidebands squeezing measurement (blue), oscilloscope for the pulse-by-pulse measurement (orange).}
\label{sqzcurve}
\end{figure}

Using this procedure, we performed a pulse-resolved (pulse-by-pulse) squeezing and  antisqueezing measurement in seven HG modes (up to HG$_6$). The measured values are comparable with the non-pulse-resolved sideband squeezing measurements performed with the spectrum analyzer of Table~\ref{tab:sqz}, as shown in Fig.~\ref{sqzcurve}
The small discrepancies observed between the two data sets can be explained by the lower signal-to-noise ratio in the pulse-by-pulse measurement.


We hence revealed the same multimode structure via the pulse-by-pulse measurement and the sidebands measurement. 
The results confirm that the multimode spectral structure exists at a single pulse level; this opens the way to the generation of entangled structures at the repetition rate of the laser. 
It is also possible to use delay lines or multiplex multiple waveguided sources \cite{Asavanant19,Larsen19,Madsen2022} to engineer more elaborate entangled structures from light pulses and spectral modes, as suggested in Fig. \ref{scheme}.


\paragraph*{\textbf{Covariance matrix and entanglement}}

In the previous sections, we presented the characterization of the down-converted light in the Hermite-Gauss basis of optical frequencies, where the system is described as a collection of independently squeezed states. However, it is also possible to measure this state in other bases, revealing the presence of entanglement between the modes of a chosen basis.
Here, we show the reconstruction of the state covariance matrix in the frequency-band basis, measuring amplitude and phase noise correlations between different frequency bands of the output light spectrum. 

\begin{figure}
 \includegraphics[width=0.4\textwidth]{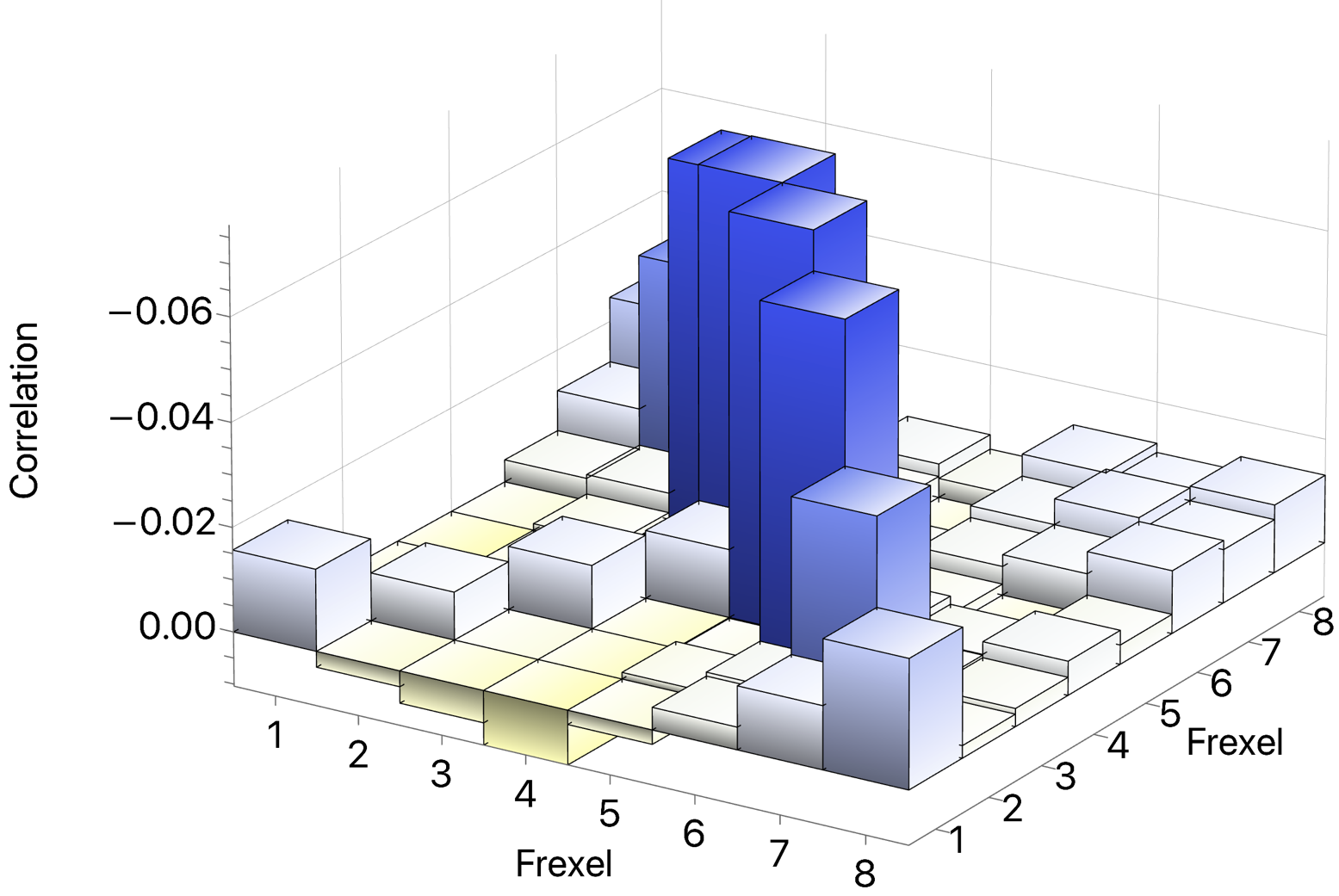}\\
 \includegraphics[width=0.4\textwidth]{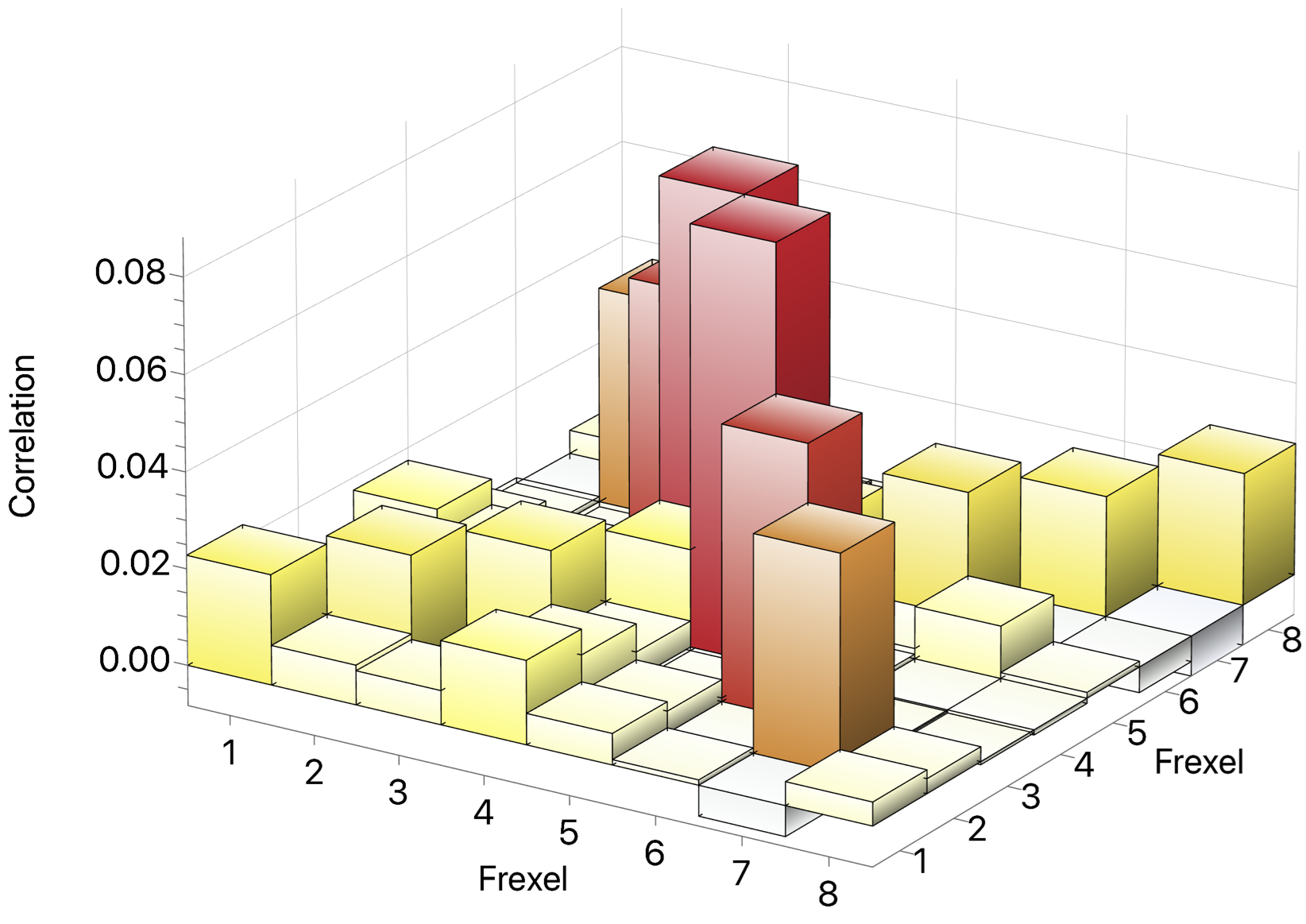}
\caption{Experimental matrices $\mathbf{V}_{qq}-\mathbbold{1}$ (upper) and $\mathbf{V}_{pp}-\mathbbold{1}$ (lower) that show the correlations among the 8 frexels, where the identity has been subtracted for better visibility of the off-diagonal elements. The z-axis of the upper plot is reversed to properly show the anti-diagonal. Strong correlations emerge in the anti-diagonal of both matrices, a characteristic of the type 0 SPDC.}
\label{fig:covariancematrix}
\end{figure}

For that, we divide the LO spectrum in 8 individual frequency bands of 5 nm, that we call \textit{frexels}. If we call $q_i$ and $p_i$ the quadratures associated to each of these frexel modes, the covariance matrix is defined as $V_{x_i,x_j} = (\langle \delta x_i \delta x_j\rangle + \langle \delta x_j \delta x_i\rangle)/2$ where $x$ is either $q$ or $p$ and $\delta x = x - \langle x \rangle$ . We measure it using the pulse shaper to address the spectrum of each frexel as well as all combinations of two frexels. This leads to the measurement of 36 homodyne traces.
The reconstructed covariance matrices can be seen in Fig.~\ref{fig:covariancematrix} (see Methods for details). The diagonal elements show the noise in each frequency band, while the off-diagonal terms reveal the correlations between the different frequency bands. The strong correlations observed in the antidiagonal are typical of the joint spectral structure of the generated photon-pair in the chosen type-0 SPDC.

The off-diagonal correlations can be used to evaluate entanglement between any bipartition of the 8 frequency bands~\cite{Gerke15}. We have measured entanglement in 114 of the  127 possible bipartitions.
The amount of entanglement, measured via a positive partial transpose criterion, varies along the different bipartitions. It indeed reflects 
the structure of the narrow type-0 joint-spectrum of the entangled photon-pairs  generated by the non-linear process into the waveguide. The quantum correlations, as also shown in the covariance matrix, are mostly shared between frequency bands symmetrically spaced from the central optical frequency, i.e. 4-5, 3-6 and 2-7. For such reason bipartitions that do not decouple these pairs of frexel are not guaranteed to be entangled.
Details can be found in Methods. 

\begin{figure}
\includegraphics[width=0.5\textwidth]{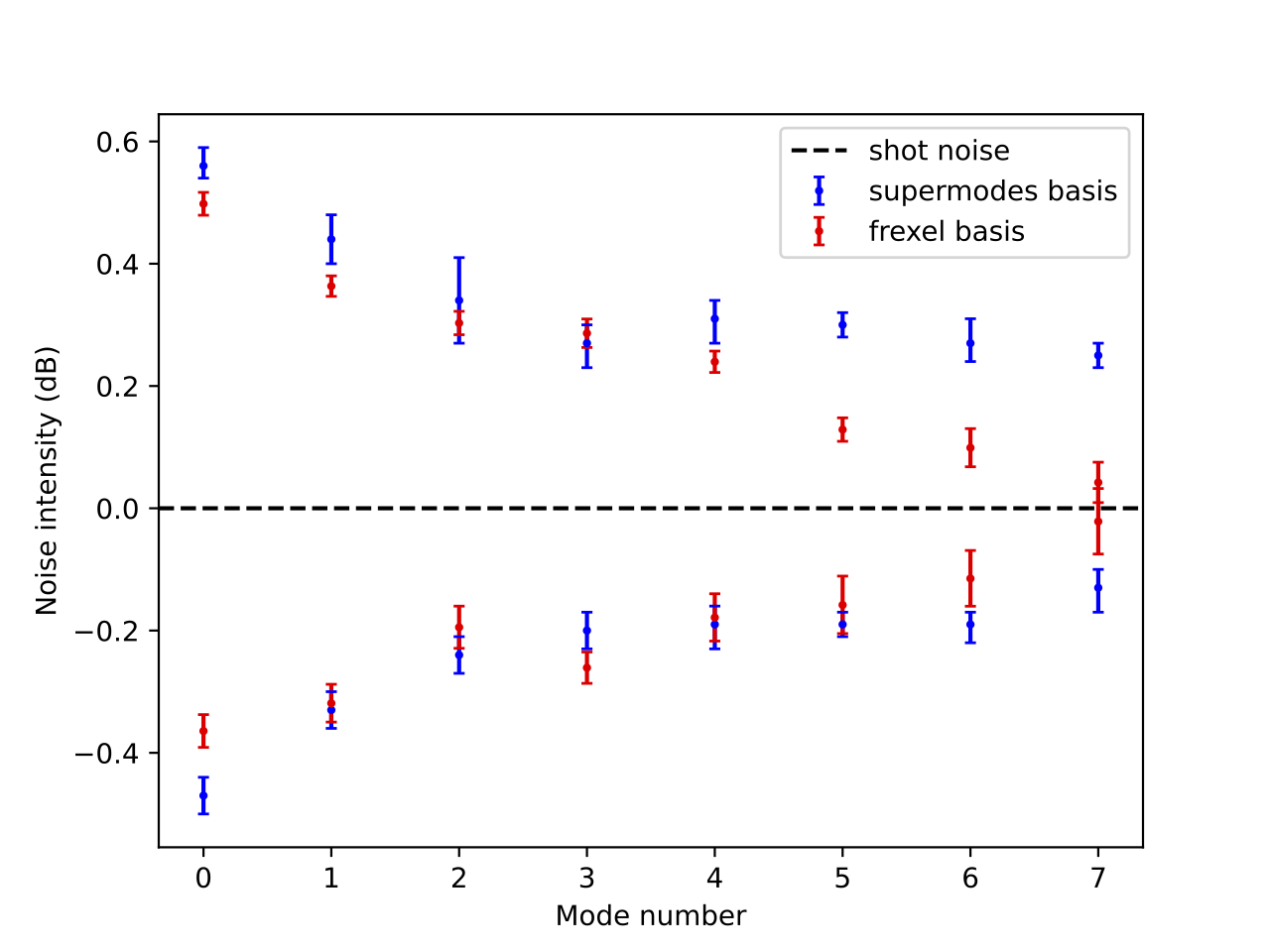}
\caption{Comparison of the squeezing and antisqueezing values measured in the supermodes Hermite-Gauss basis (blue) with those obtained from the reconstruction of the covariance matrix, measured in the frexels basis (red). The two measurement sets share the same measurement technique (non-time resolved sideband squeezing measurement) and device (spectrum analyser) but different measurement basis.}
\label{eigenvalues_error}
\end{figure}

From the reconstructed covariance matrix, we can also recover the squeezing values associated with the uncorrelated squeezed modes of the system, i.e., the supermodes. To do so, we diagonalize the covariance matrix. The results are shown in Fig. \ref{eigenvalues_error}, where
the diagonal elements are compared with the squeezing values shown in 
Fig. \ref{sqzcurve}. Small deviations are due to the larger resolution (larger signal-to-noise ratio) of the measurement in the supermode basis, compared to the 8-pixel basis.

We have shown that the basis change, via mode-selective homodyne detection, can be used to build/modify entanglement correlation between optical frequency modes. This is an essential precondition to build the networks of entangled modes shown in the lower right part of Fig. \ref{scheme}. 

\section{Discussion}
We demonstrated a single-pass source producing 21 independently squeezed spectral modes at 156 MHz. This allows the simultaneous multiplexing of the temporal and (optical) spectral degrees of freedom of deterministic quantum resources. 
We produced the largest number of spectral modes of squeezed vacuum measured via homodyne detection in a single-pass configuration. Tailoring of the phase-matching condition and the pump spectrum can be used to increase the amount of squeezing per spectral-mode. We can in fact tolerate the consequent decreasing of number of squeezed modes, that are predicted to be 98 by the Schmidt number in the actual configuration. A number that we cannot fully access in any case, given the large spectral span of the high-order modes.
Nonetheless, the main result of this work  is the simultaneous multiplexing of both degrees of freedom; this constitutes a versatile building block of the 3-dimensional structures that are required to build a scalable quantum computer \cite{Bourassa2021}. 
Large 2-dimensional CV cluster  states \cite{Asavanant19,Larsen19} have been generated by combining time modes from multiple squeezing sources, a 3-dimensional reconfigurable structure obtained via time multiplexing of a squeezing source have been recently exploited in a boson sampling setting \cite{Madsen2022}. Here, we demonstrate a resource that, by offering multiple squeezed modes at the time of a  given pulses, can be  shaped as a 3-dimensional entangled structure (as shown in the scheme of Fig. \ref{scheme}). The reconfigurability in our scheme is offered by multiplexing via spectrally resolved homodyne detection  \cite{Ansquer21}. So different 2-dimensional spectral structure can be concatenated in the time domain as fast as the repetition rate of the pulsed laser source (156 MHz). Moreover, spectrally mode-selective non-Gaussian operations \cite{Ra20}, needed for quantum computing protocols \cite{Bourassa2021,Eaton2022}, can be implemented on selected nodes of the 3-D structure.

\begin{figure*}
\includegraphics[width=1\textwidth]{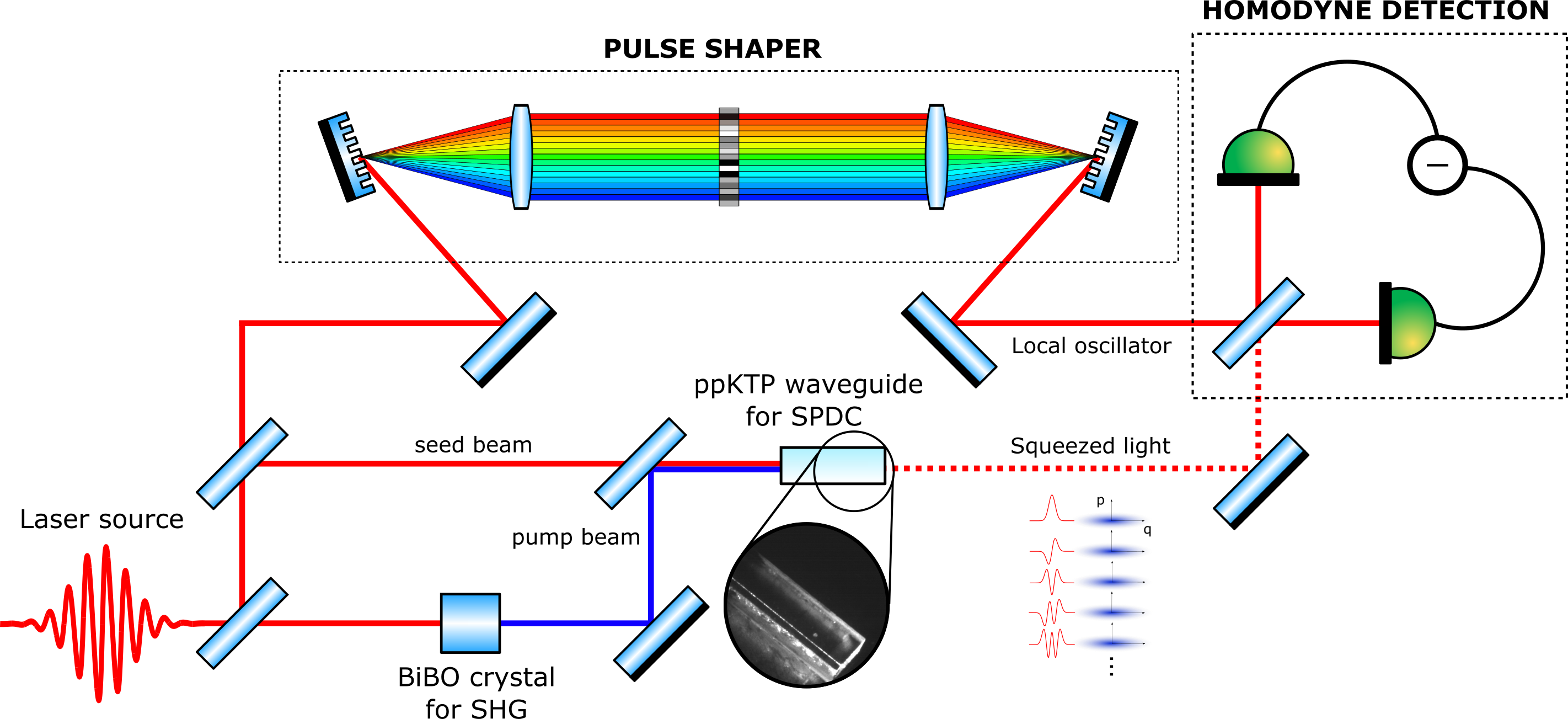}
\caption{Experimental setup for the generation of multimode squeezed vacuum multiplexed in time and frequency. The laser source produces a train of 22-fs pulses centered at 795 nm at 156 MHz. The light undergoes up-conversion (via BiBO crystal) and then pumps a ppKTP non-linear waveguide. The down-converted output is measured by homodyne-detection and the spectral mode to be measured is selected by shaping the LO using a pulse-shaper. }
\label{exp_setup}
\end{figure*}

\section{Methods}
\paragraph*{\textbf{Experimental setup}}
The laser source is a femtosecond mode-locked Titanium-Sapphire (Ti:Sa) (Femtosource Synergy+ from Spectra-Physics), delivering trains of 22-fs pulses at a repetition rate of $f_{rep} =$ 156 MHz ($\tau_{rep} \simeq$ 6 ns). The corresponding optical spectrum is  centered at 795 nm with a  FWHM of about about 42 nm. 
At the laser output, the average power is 1W. The IR beam is 
split in three beams: 30\% is sent to a pulse shaper, and serves as a reference for the homodyne detection, 10\% is used to seed the SPDC, and the final 60\% is used to pump a 1-mm long BiBO (Bismuth Triborate) crystal, critically phase-matched for second-harmonic generation. After the SHG, the pulsed source is centered at 397.5 nm and has a 
spectrum of 0.7 nm (FWHM). 
The up-converted light pumps a periodically poled KTP (Potassium Titanyl Phosphate) waveguide, and undergoes a collinear type-0 SPDC process generating a train of quadrature-squeezed pulses. The temperature of the waveguide is set to its value of optimal conversion efficiency (89.1 $^{\circ}$C).
The pump and seed beams are coupled inside the waveguide with an achromatic lens 
 while the exiting beams are collimated with an IR-coated lens 
 and the residual pump light is filtered out with a dichroic mirror. Inside the waveguide, the temporal matching of the seed and the pump pulses is achieved by optimizing the phase-sensitive parametric amplification to $\sim$20\%. 

The quantum state generated by the parametric process is characterized via homodyne detection. To unveil the multimode structure of the squeezed light, we use a pulse shaper to modify the LO spectrum according to the spectral mode that we want to measure. The overall transmission of the setup is 70\%, and it can also be used to correct for undesired spectral dispersion. 

The shaped LO is interfered on  a balanced beam splitter with the seed beam first, for alignment, then with the signal. Their relative phase is scanned with a piezo-electric actuator placed in the LO path.  
The visibility between  of the interference fringes of the LO and the seed beam is maximized to reach the equivalent mode-matching efficiency $\eta_{mm} = 90\%$, a value that accounts for both the temporal and spatial overlap, while the spectral overlap between the signal and the LO is optimized by the pulse-shaping.
A scheme of the experiment is shown in Fig.~\ref{exp_setup}.

\paragraph*{\textbf{Reconstruct the covariance matrix}}

To reconstruct the state, we measure the amplitude and phase   correlations. Assuming that the pump is real, the quadratures $q$ and $p$ are uncorrelated, thus the covariance matrix $\mathbf{V}$ takes a block-diagonal form:
\begin{equation}
\mathbf{V}= \begin{pmatrix} \mathbf{V}_{qq} & 0 \\ 0 & \mathbf{V}_{pp}\end{pmatrix}\\
\end{equation}
where 
\begin{equation}
\mathbf{V}_{xx}= \begin{pmatrix} \Delta^2 x_1& \langle  x_1x_2\rangle  &\dots \\ \langle x_1 x_2\rangle  & \Delta^2x_2 & \\ \vdots  &  & \ddots \\\end{pmatrix}\\
\end{equation}
with $\boldsymbol{x}=\{\mathbf{q},\mathbf{p}\}$. 
We denote $i+j$ the mode combination of the frexels $i$ and $j$ and we define its quadrature operators as $\hat{x}_{i+j}=(\hat{x}_i+\hat{x}_j)/\sqrt{2}$, provided that the optical power detected by the frexels $i$ and $j$ is the same. Given that the mean value of the quadrature is $\langle q \rangle = \langle p \rangle =0$, we compute the off-diagonal elements with the formula:
\begin{equation}
\langle x_i x_j\rangle = \left[ \Delta^2x_{i+j} - \frac{\Delta^2 x_i+\Delta^2 x_j}{2}\right]
\label{eq:xixj}
\end{equation}
Each right-hand term of equation \ref{eq:xixj} is measured with the spectrum analyser, where they correspond to the minima (for the $q$ quadrature) or the maxima (for the $p$ quadrature) of the homodyne traces.
To recover the uncorrelated squeezed modes of the system, i.e., the supermodes, the covariance matrix needs to be diagonalized. To do so, we use a Gram-Schmidt process to orthogonalize the eigenmodes associated to the antisqueezed modes, and we use them to construct the matrix applied to the covariance matrix to operate the basis change. 
The relative sign between the eigenvalues of $\mathbf{V}_{qq}$ (equivalently $\mathbf{V}_{pp}$) stands for the relative phase between the squeezing ellipses of the different supermodes. 
The uncertainty of the squeezing values stems from the uncertainty of the squeezing or anti-squeezing peaks measured with the spectrum analyzer.

\paragraph*{\textbf{Entanglement between different bipartitions}}

\begin{figure}
\includegraphics[width=0.5\textwidth]{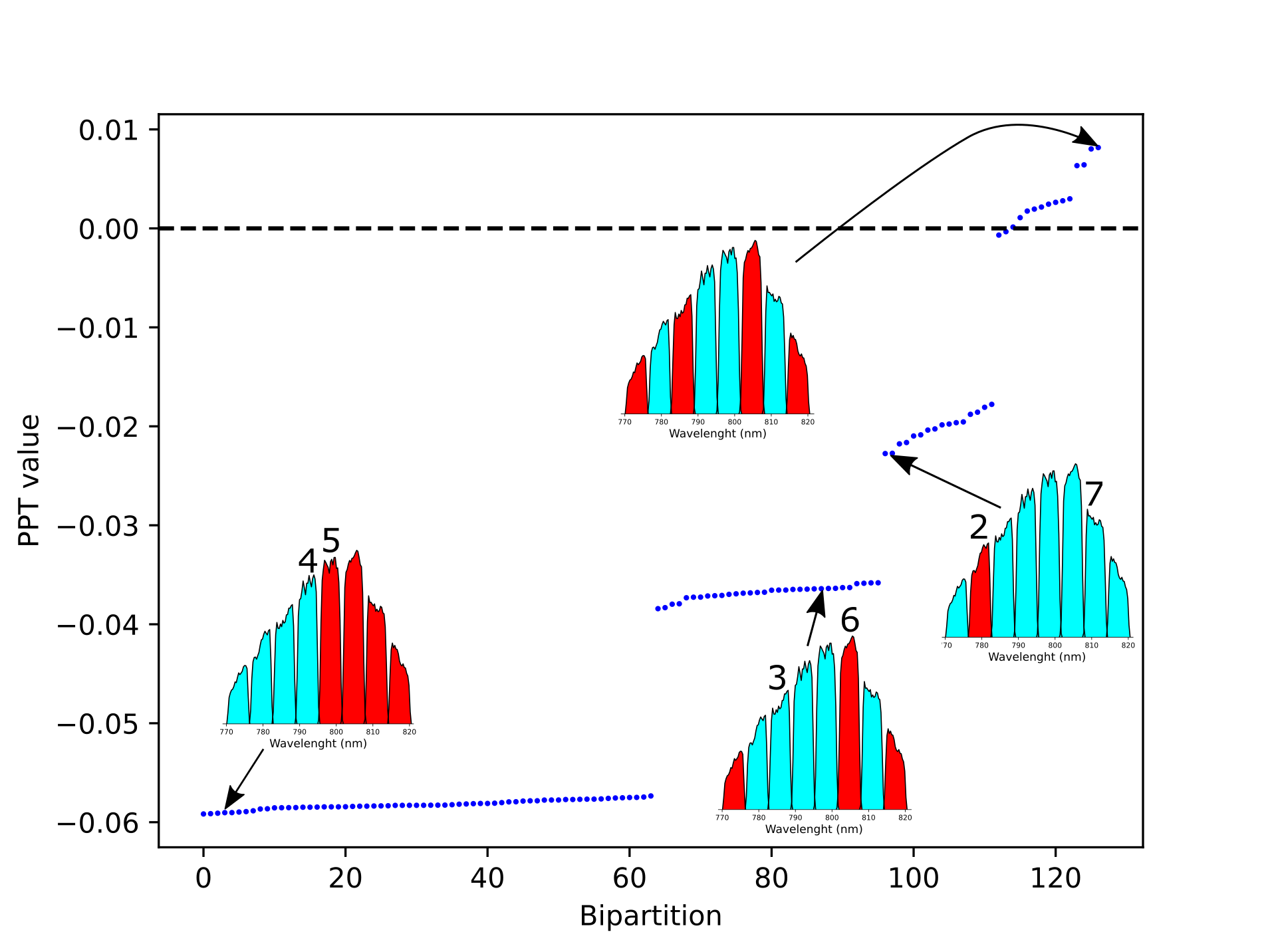}
\caption{PPT inseparability value (lowest eigenvalue of the PPT matrix) for all of the 127 possible bipartitions of 8 frexels, ordered from the smallest to the highest. All the bipartitions below the dashed line (negative PPT value) are entangled. 4 distinct bands of values are distinguishable, depending on which pair of frexels are separated into two different bipartitions (4-5, 3-6, 2-7 or none of them).}
\label{fig:PPTcriterion}
\end{figure}

We can assess bipartite entanglement between two given bipartitions of the state with the positive partial transpose criterion (PPT) for CV~\cite{Simon00}. We measure the PPT value for each of the possible 127 bipartitions, i.e. different ways to divide the 8 frequency bands into two sets. This is done by carrying out a partial transpose operation on the density operator of the state, that maps its covariance matrix $\mathbf{V}$ into  $\tilde{\mathbf{V}} = \mathbf{\Lambda}\mathbf{V}\mathbf{\Lambda} $, where $\mathbf{\Lambda}$ is the operation that changes the sign of the $p$ coordinate of one of the two bipartitions. A necessary condition for separability of the state reads $\mathbf{P}=\tilde{\mathbf{V}}-i\mathbf{J} \geq 0$, where $\mathbf{J}$ is the $2n \times 2n$ symplectic form 
\[
J = \begin{pmatrix}  0 & -\mathbbold{1} \\ \mathbbold{1} & 0\end{pmatrix}
\]
If the above condition is not fulfilled, the two bipartitions are entangled. The smallest eigenvalue of $\mathbf{P}$ (PPT value) for each possible bipartition is shown in Fig.~\ref{fig:PPTcriterion}, where we see that inseparability holds for ~90\% of the bipartitions. Moreover, we can distinguish 4 different bands of PPT values: a lower, a lower-intermediate, an upper-intermediate and an upper band. The lower band is the band with the highest degree of entanglement (lower PPT value) and it
corresponds to a partitioning choice where frexels 4 and 5 are separated in two different bipartitions. The lower-intermediate band contains bipartitions that separate frexel 3 and 6 (but 4 and 5 are in the same one), while the upper-intermediate sees frexels 2 and 7 in separate bipartitions (but 4-5 and 3-6 are in the same). Finally, the upper band is mostly composed of non-entangled bipartitions, where the above mentioned pairs of frexels are kept coupled together and not split into two different bipartitions. The most external frequency bands of the spectrum (frexels 0 and 7) have a less significant role. 

\paragraph*{\textbf{Measuring squeezing pulse-by-pulse}}
Pulse-by-pulse squeezing measurement in the temporal domain is performed using a fast oscilloscope, which allows to directly measure quantum fluctuations of the homodyne signal. It requires careful alignment of the homodyne detector, which is achieved by precisely distributing the optical power applied to each photodiode (2.5 mW at most), maximizing the temporal overlap between input pulses and adjusting the bias voltage applied to the photodiodes. These steps are followed to minimize the CMRR (Common Mode Rejection Ratio) at 156 MHz.

For each squeezing measurement we collect two million data samples every 50 $\mu$s. This corresponds to an average of 7.8 million pulses per squeezing value. We use the signal from one of the photodiodes as a reference to determine the time window corresponding to a single pulse. Additionally, we verify that the cross correlation function is minimized to confirm that the measured value corresponds unequivocally to the average squeezing of a single pulse. The phase of the LO is varied with a piezoelectric actuator driven much slower than the measurement rate of the oscilloscope. 
The spectrum analyzer is synchronized with the oscilloscope and we measure the quadrature variance for different positions of the piezoelectric actuator. 
We keep the squeezed and antisqueezed quadratures, discard intermediate values, and measure the variance of the shot noise (not squeezed) to set a reference for each measurement.


\paragraph*{\textbf{Design and characterization of the nonlinear waveguide}}
Periodic poling is used to increase the gain of the SPDC by compensating for the phase mismatch (quasi-phase-matching). The poling period depends on the crystal properties and on the wavelength used in the nonlinear process. For this experiment, we used numerical simulations to determine the poling period ($\Lambda = 3.19 \, \mu \text{m}$) and the interaction length ($\ell= 1$ mm) of the nonlinear waveguide. The company AdvR Inc. designed and manufactured the waveguide chip. It holds 30 graded-index waveguides embedded in a \textcolor{teal}{x}-cut PPKTP crystal. The chip is 5.7-mm long to guarantee the quality of the input and output facets of the crystal, and each waveguide is made of a periodically-poled portion $\ell$, where the parametric process occurs, followed by a non-poled (not phase-matched) portion.

The chip manufacturing process does not guarantee the production of identical waveguides \cite{Bierlein1987}. Therefore, we characterize them all by measuring the coupling efficiency and the output spectrum and we select the best ones for our experiment. The waveguides are designed to be single mode at 795 nm but, in practice, we observed the propagation of higher-order spatial modes. We believe this is caused by the gradient of refractive index in the vertical direction of the waveguide. Adjusting the waveguide height ensures that only the fundamental mode at 795 nm can couple inside the waveguide. Additionally, we select the waveguides whose output spatial profile provides a good visibility with the LO. Ultimately, we chose a 3-$\mu$m-wide waveguide with a coupling efficiency of 57\%.

At lower wavelengths, the waveguides are spatially multimode  but only the fundamental mode is optimal for the SPDC. Thus, spatial mode-matching is required to shape the pump and optimize the coupling of the fundamental mode compared to higher-order modes.


\paragraph*{\textbf{Wideband balanced homodyne detector}}
In this section, we summarize the features of a homemade wideband homodyne detector. The electrical circuit uses two Si PIN photodiodes from Hamamatsu with 300 MHz bandwidth and 91 \% quantum efficiency at 795 nm. After detection, the photocurrents generated from both photodiodes are subtracted and amplified to generate the homodyne signal that we measured and analyzed with an Agilent 86142B spectrum analyzer and a Teledyne Waverunner oscilloscope. We measured a CMRR of 64 dB at 300 $\mu$W. More details on the design of the detector can be found in \cite{Kouadou2021}


\section{Acknowledgements}
We thank B\'ereng\`ere Argence for her support in the design and realisation of the homodyne detector.
This   work   was   supported   by   the   European   Research Council under the Consolidator Grant COQCOoN (Grant No.  820079) and by Region Ile-de-France in the framework of DIM SIRTEQ.


\bibliography{biblioMus}


\end{document}